\documentclass[aps,prb,twocolumn,groupedaddress,showpacs,floatfix]{revtex4}
\usepackage{graphicx}

\begin{document}

\title{Dimensional-scaling estimate of the energy of a large system from 
that of its building blocks: Hubbard model and Fermi liquid}
\author{K. Capelle and L. N. Oliveira}
\email{capelle@if.sc.usp.br}
\affiliation{Departamento de F\'{\i}sica e Inform\'atica\\
Instituto de F\'{\i}sica de S\~ao Carlos\\
Universidade de S\~ao Paulo\\
Caixa Postal 369, S\~ao Carlos, 13560-970 SP, Brazil}
\date{\today}

\begin{abstract}
A simple, physically motivated, scaling hypothesis, which becomes exact in 
important limits, yields estimates for the ground-state energy of large, 
composed, systems in terms of the ground-state energy of its building blocks. 
The concept is illustrated for the electron liquid, and the Hubbard model. 
By means of this scaling argument the energy of the one-dimensional
half-filled Hubbard model is estimated from that of a 2-site Hubbard dimer,
obtaining quantitative agreement with the exact one-dimensional Bethe-Ansatz
solution, and the energies of the two- and three-dimensional half-filled 
Hubbard models are estimated from the one-dimensional energy, recovering 
exact results for $U\to 0$ and $U\to \infty $ and coming close to Quantum 
Monte Carlo data for intermediate $U$.
\end{abstract}



\maketitle

\newcommand{\be}{\begin{equation}}
\newcommand{\ee}{\end{equation}}
\newcommand{\bea}{\begin{eqnarray}}
\newcommand{\eea}{\end{eqnarray}}
\newcommand{\bi}{\bibitem}

\renewcommand{\r}{({\bf r})}
\newcommand{\rp}{({\bf r'})}

\newcommand{\ua}{\uparrow}
\newcommand{\da}{\downarrow}
\newcommand{\la}{\langle}
\newcommand{\ra}{\rangle}
\newcommand{\dg}{\dagger}

Scaling techniques have propped up research since the early days of
modern science. In the traditional version, which we call {\em
discrete scaling}, one measures the properties of a body of diameter
$D$ to determine those of a similar body of diameter $D'$, typically
much larger than $D$. This approach is widely employed, e.g., in
fluid dynamics. A more recent approach, extensively explored
in the theory of critical phenomena, focuses on the relation between
differential increments in the size $L$ of a complex system and the
resulting changes in its physical properties. To emphasize that $L$ is
a continuous variable, we refer to this procedure as {\em continuous
scaling}.
Continuous scaling finds a parallel in studies, also pioneered in the
theory of critical phenomena, that treat the dimensionality $d$ of a
system as a continuous variable, and hence will be called {\em
continuous dimensional scaling}. A classic illustration is the
$\epsilon$ expansion, which extracts the properties of
$(4-\epsilon)$-dimensional systems from the exact mean-field solution of
the $4$-dimensional Gaussian model.\cite{kogut} Other recent examples are
found in dynamical mean-field theory\cite{dmft} (DMFT) and in
density-functional theory (DFT) for the Heisenberg model.\cite{hemo}
                                                                              
Short as it is, this overview points to an as yet unprobed form of scaling:
{\em discrete dimensional scaling}. We are unaware of attempts to extract
the properties of a system of dimensionality $d$ from the known properties of
the same system in a different dimensionality. In the present work we
explore this possibility. As a specific nontrivial model, to illustrate the 
general idea of discrete dimensional scaling, we choose the Hubbard model.

The Hubbard model was originally proposed on a three-dimensional lattice
for the description of correlations in narrow-band materials.\cite{hubbard}
Its two-dimensional version is often considered the correct minimal model for 
describing the cupper-oxide planes in cuprate superconductors.\cite{anderson} 
The one-dimensional Hubbard model has received much attention due to the 
Luttinger liquid and Mott insulator phases it displays.\cite{schlottmann} 
In the limit of infinite number of dimensions the Hubbard 
model was at the beginning of DMFT.\cite{dmft} 
In all dimensions, the model serves as a laboratory and prime test case for 
our understanding of many-body physics and, in particular, strong Coulomb 
correlations. In standard notation and for any dimensionality $d$, the model 
reads
\be
\hat{H}=\hat{T}+\hat{U}=
-t\sum_{\langle ij \rangle,\sigma} c_{i\sigma}^\dagger c_{j\sigma}
+U\sum_i c_{i\ua}^\dagger c_{i\ua}c_{i\da}^\dagger c_{i\da},
\label{1dhm}
\label{hm}
\ee
where $t$ and $U$ are the kinetic energy and interaction parameters,
respectively, and $\langle ij \rangle$ denotes a sum over nearest neighbors
on a $d$-dimensional lattice.

The one-dimensional homogeneous Hubbard model has an exact solution in terms 
of the Bethe Ansatz,\cite{schlottmann,liebwu} one of the key results of which
is a closed expression for the per-site ground-state energy at half filling, 
as a function of $U$, 
\be
\frac{E(U,d=1)}{t N_s}= -4\int _0^\infty dx
\frac{J_0(x)J_1(x)}{x[1+\exp(Ux/2)]},
\label{lw}
\ee
where $N_s$ is the number of sites and 
$J_0$ and $J_1$ are zero and first order Bessel functions. Such simple 
expressions for the total energy of nontrivial models are scarce in many-body 
physics, and where they exist they provide, in addition to considerable
algebraic simplifications, insights into the nature of the model that would
be hard to obtain from numerical data on its own. 

The purpose of the present paper is to investigate to which extent knowledge
aquired in low-dimensional situations, such as Eq.~(\ref{lw}), can be useful 
in higher-dimensional systems, where exact solutions are not available. 
Mathematically, low and higher dimensional systems are solutions of the same 
type of differential equation, Schr\"odinger's equation. Physically, higher 
dimensional systems are built out of lower-dimensional building blocks. Both 
of these facts suggest that higher-dimensional systems cannot be arbitrarily 
different from lower-dimensional ones. Mathematical and structural constraints 
imply the existence of (perhaps complicated) connections between quantities
pertaining to one and to higher-dimensional systems. Here we identify and
explore such constraints.

A first illustration of the concept of dimensional scaling is provided by
bound homogeneous Coulomb systems, such as the Fermi liquid, described by the 
Hamiltonian
\be
\hat{H}_{Clb}=-\frac{1}{2}\sum_i \nabla_i^2 + \sum_{i>j} \frac{1}{r_{ij}} +W,
\label{liquid}
\ee
where $W$ represents the constant potential energy of the uniform charge 
background, or of confining walls. For such a system satisfaction of the 
quantum-mechanical virial theorem \cite{lali}
guarantees dynamical stability of the 
ground state. This theorem states that the total ground-state energy 
(comprising kinetic, potential and interaction energy) is the negative of 
the kinetic energy, i.e.,\cite{footnote0} $E_{tot}=-E_{kin}$. Any relation
of the form $E_{tot}=A E_{kin}$, with constant $A$, allows to write the 
identity $E_{tot}(d)/E_{tot}(d') = E_{kin}(d)/E_{kin}(d')$, which implies 
that the $d$-dimensional total energy can be obtained by scaling the 
$d'$-dimensional total energy according to 
\be
E_{tot}(d) = E_{kin}(d) \frac{E_{tot}(d')}{E_{kin}(d')}.
\label{virial}
\ee
Equation (\ref{virial}) is a first example of dimensional scaling, showing 
that the solution of a low-dimensional problem can generate a solution to a 
higher-dimensional problem of the same type. The virial theorem plays the role 
of the constraint mentioned above.

Next, we explore the question if similar expressions can also be obtained for 
the Hubbard model, whose total ground-state (GS) energy we denote by $E(U,d)$. 
Instead of the virial theorem, whose above form does not apply to the 
Hubbard model, we here base our analysis on the Hellman-Feynman theorem, 
which allows to cast the kinetic energy $E_{kin}(U,t)=E(U,t)-E_{int}(U,d)$ as 
\be
E_{kin}(U,t)=t \frac{\partial E(U,t)}{\partial t} 
= E(U,t)-U \frac{\partial E(U,t)}{\partial U}.
\label{hftheorem}
\ee
The second term on the right-hand side vanishes both at small $U$, where the 
total ground-state energy $E(U,d)$ is linear in $U$, and at large $U$, where 
$E(U,d) \propto 1/U$. Hence, in both limits $E(U,d)=+E_{kin}(U,d)$. As 
for the electron liquid, this equality gives rise to the dimensional-scaling 
expression for the $d$-dimensional ground-state energy,
\be
E(U,d) \approx
E^{DS}(U,d) = E_{kin}(U,d) \frac{E(U,d')}{E_{kin}(U,d')}
\label{linear}
\ee
but unlike for the electron liquid, this is exact only in two limits, and 
approximate in between. Its use at all $U$ embodies a linear 
dimensional-scaling hypothesis, whose validity must be carefully checked. 
Our first check is an application 
of Eq.~(\ref{linear}) to one-dimensional Hubbard chains, whose energies we try 
to recover by means of Eq.~(\ref{linear}) from that of a zero-dimensional 
Hubbard model. A truly zero-dimensional model is useless for the present 
purpose, but a two-site system is as close to zero dimensionality as one can 
get without losing either of the two defining terms of the Hubbard model, the 
on-site interaction and the nearest-neighbour hopping.

\begin{figure}
\includegraphics[height=80mm,width=65mm,angle=-90]{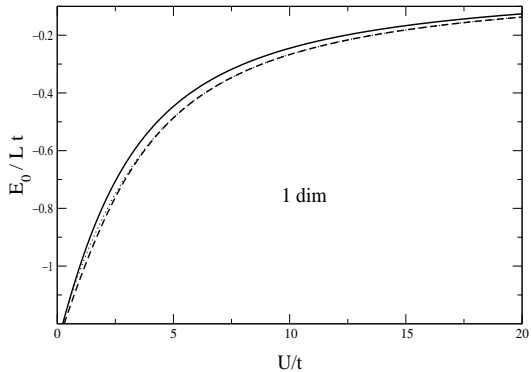}
\caption{\label{fig1}
Ground-state energy of the one-dimensional half-filled Hubbard model at
positive $U$.
Lower (dashed) curve: exact result, Eq.~(\ref{lw}).
Middle (dotted) curve:  present expression (\ref{1dscaling}), representing
linear dimensional scaling with interacting kinetic energies.
Upper (full) curve: present expression (\ref{1dnonint}), representing
linear dimensional scaling with noninteracting kinetic energies.}
\end{figure}

Applied to this case, Eq. (\ref{linear}) reads
\be
E(U,1) \approx 
E^{DS}(U,1) = E_{kin}(U,1) \frac{E(U,L=2)}{E_{kin}(U,L=2)},
\label{1dscaling}
\ee
where $L$ is the number of sites.
$E(U,1)$ is exactly known from Eq.~(\ref{lw}),
which thus allows a stringent test of the concept of dimensional scaling.
The two-site Hubbard model can be diagonalized analytically for any $U$. For 
a half-filled band the exact per-site ground-state energy is
\be
E(U,L=2)=\frac{U}{4}-\sqrt{\frac{U^2}{16}+t^2}.
\label{2site}
\ee
The kinetic energies in Eq.~(\ref{1dscaling}) can be obtained from 
(\ref{lw}) and (\ref{2site}) via
$E_{kin}=t \partial E/\partial t$. All ingredients on
the right-hand side of (\ref{1dscaling}) are thus known, and the resulting 
prediction for $E(U,1)$ can be compared to the exact result for the infinite 
chain. This comparison, in Fig. \ref{fig1}, shows that linear dimensional 
scaling (dotted curve) almost coincides with the exact energy (dashed curve),
illustrating that Eq.~(\ref{1dscaling}) with (\ref{2site}) is a near-exact 
representation of the true ground-state energy, even at intermediate $U$.

In more complicated cases the exact kinetic energies may not be known. We
therefore also consider an approximation to Eq.~(\ref{1dscaling}) in which 
noninteracting kinetic energies are used in the scaling relation. For the
dimer and the chain the noninteracting per-site kinetic energies are 
$E(U=0,L=2)=-t$ and $E(U=0,d=1)=-4t/\pi$, respectively. Hence, the 
noninteracting dimensional-scaling prediction for the per-site energy of 
the infinite chain is
\be
E^{DS-0}(U,1) 
= E(0,1) \frac{E(U,L=2)}{E(0,L=2)}
= \frac{U -\sqrt{U^2+16 t^2}}{\pi}.
\label{1dnonint}
\ee
We note that in the limits of very weak and very strong interactions this
equation predicts $E^{DS}(U\to 0,1) = -4t/\pi$ and $E^{DS}(U\to \infty,1)=0$,
respectively. These are the same values also obtained from the Bethe-Ansatz
expression (\ref{lw}), showing that in spite of the additional approximation 
entailed by using the noninteracting kinetic energy ratio instead of the 
interacting one, the dimensional scaling Ansatz is still exact for very weak
and very strong interactions. For intermediate $U$, a numerical comparison 
between the exact Bethe-Ansatz result and the dimensional scaling estimate
(\ref{1dnonint}) is presented in Fig.~\ref{fig1} (full curve). 

Figure \ref{fig1} shows that, in spite of its simplicity, the dimensional 
scaling expression (\ref{1dscaling}) reproduces the function $E(U,d=1)$ almost 
exactly, while (\ref{1dnonint}) provides a very reasonable approximation to it. 
Hence, the ground-state energy of the infinite chain is quantitatively 
recovered by suitable scaling of a simple dimer calculation. Crucially, this 
scaling involves an energy ratio, and not a length ratio.

After this exploration of the concept of dimensional scaling in situations 
in which the exact result is known in closed form, we now turn to a situation
in which the higher-dimensional result is known only numerically. Namely, 
we attempt to predict the
ground-state energy of the half-filled two-dimensional repulsive Hubbard model
by applying virial-based dimensional scaling to the Lieb-Wu Bethe-Ansatz
expression (\ref{lw}). This expression thus ceases to be our target to
become the input on the right-hand side of Eq.~(\ref{linear}).

The $d$-dimensional kinetic energy $E_{kin}(d)$ of the Hubbard model is 
already an in general unknown function of $d$, so that Eq. (\ref{linear})
cannot be used directly. We can, however, still employ noninteracting 
kinetic energies in the scaling factor.
Equation (\ref{linear}) now reads
\be
E(U,2) \approx
E^{DS-0}(U,2) = E(0,2) \frac{E(U,1)}{E(0,1)}.
\label{2dscaling}
\ee
Note that all quantities appearing on the right-hand-side of (\ref{2dscaling}) 
are known: $E(U,1)$ is given by the Lieb-Wu formula (\ref{lw}), while $E(0,d)$ 
has the values\cite{shiba,vignale,hasegawa}
$E(0,1)/t N_s=-1.2732$, $E(0,2)/t N_s= -1.6211$, and $E(0,3)/t N_s= -2.0048$.
As before, Eq.~(\ref{2dscaling}), which is by 
construction exact in $d=1$ for any 
$U$, and at $U=0$ in any $d$, is also already exact in any $d$ in the limit
$U\to \infty$: it follows from Eq.~(\ref{lw}) that $E(U\to\infty,1)=0$, and 
hence $E^{LDS}(U\to \infty,d)=0$, which is the correct $U\to \infty$ limit 
for the half-filled Hubbard model in any $d$. Dimensional scaling with
noninteracting kinetic energies is thus still exact both for very weak and 
very strong correlations.

\begin{figure}
\includegraphics[height=80mm,width=65mm,angle=-90]{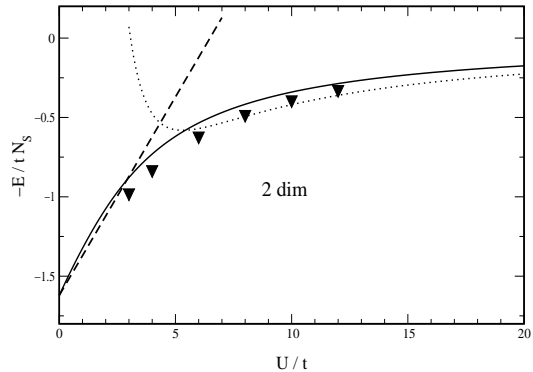}
\caption{\label{fig2}
Ground-state energy of the two-dimensional half-filled Hubbard model.
Filled triangles: VMC data extrapolated to the thermodynamic limit,
from Ref.~\onlinecite{shiba}.
Dashed curve: Hartree-Fock energy.
Dotted curve: $U\to\infty$ expansion of Ref.~\onlinecite{takahashi}.
Full curve: present expression (\ref{2dscaling}).
}
\end{figure}

A more comprehensive check on (\ref{2dscaling}) is obtained by comparing with 
precise numerical data obtained by Monte Carlo 
simulations.\cite{shiba,hasegawa,white,imada} Below we adopt as a benchmark 
the Variational Monte Carlo (VMC) data of Yokoyama and Shiba.\cite{shiba} 
Their values do not deviate substantially from those obtained
by other groups.\cite{footnote1}
Figure~\ref{fig2} compares, for the $d=2$ Hubbard model, the 
dimensional scaling expression (\ref{2dscaling}) with these benchmark data,
and with known large and small $U$ limits.

\begin{figure}
\includegraphics[height=80mm,width=65mm,angle=-90]{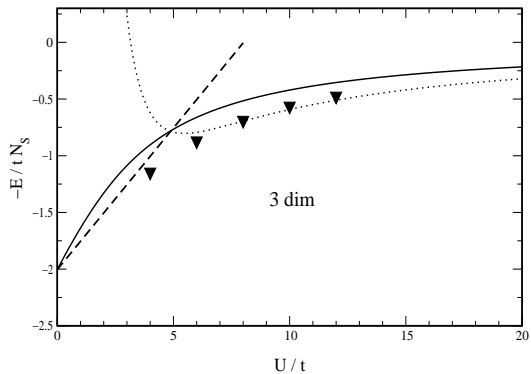}
\caption{\label{fig3}
Ground-state energy of the three-dimensional half-filled Hubbard model.
Filled triangles: VMC data on a $6\times 6 \times 6$ cluster (not
extrapolated to the thermodynamic limit) from Ref.~\onlinecite{shiba}.
Dashed curve: Hartree-Fock energy.
Dotted curve: $U\to\infty$ expansion of Ref.~\onlinecite{takahashi}.
Full curve:  present expression (\ref{linear}), representing linear
dimensional scaling.}
\end{figure}

Finally, we turn to the three-dimensional case. The 3d energies could be
constructed in terms of the 2d ones, in the same way as before, but since
the 2d energies themselves are not available in closed form this is not 
immediately practical. Instead, we subject the dimensional scaling idea
to a more severe test, and try to predict the 3d energies directly from
the 1d energies. Equation (\ref{linear}) then becomes
\be
E(U,3) \approx
E^{DS-0}(U,3) = E(0,3) \frac{E(U,1)}{E(0,1)}.
\label{3dscaling}
\ee
A comparison of the resulting curve with QMC data and
known large and small $U$ limits is shown in Fig. \ref{fig3}.

Figures \ref{fig2} and \ref{fig3} show how 1d to 2d and 1d to 3d dimensional 
scaling, as expressed by Eqs.~(\ref{2dscaling}) and (\ref{3dscaling}), merges 
into the exact expressions in the limits $U\to 0$ and $U\to \infty$, and
comes close to the QMC data at intermediate $U$. Not unexpectedly, the step 
from $d=1$ directly to $d=3$ entails a larger margin of error than those from 
$d=0$ to $d=1$ and from $d=1$ to $d=2$, because the dimensional gap to be 
bridged by a simple scaling hypothesis is larger. The concatenation of the 
$d=0$ to $d=1$ expression with the $d=1$ to $d=3$ expression shows that an 
exact solution of the two-site system, followed by a double application of 
linear dimensional scaling, is sufficient to predict energies of the 
three-dimensional system with a worst error of the order of 20 percent (and 
much less than that for most values of $U$), {\it without empirical adjustments
or fitting parameters}. Remarkably, a two-site system thus contains enough 
nontrivial many-body effects that a simple scaling hypothesis is sufficient 
to recover from it a large part of the energy of chains, planes and bulk 
systems.

In summary, we have introduced the concept of dimensional scaling of 
ground-state energies. One explicit realization of this concept, based on 
the virial theorem, is exact for the electron liquids. Another, based on the 
Hellman-Feynman theorem, allows to predict the energies of an infinite Hubbard
chain from that of a dimer with unexpected accuracy, and continues reasonable 
when extended to planes and bulk systems. Although the noninteracting 
kinetic energies in the scaling factor introduce an additional approximation,
the dimensional scaling ansatz remains exact in the limits of weak and strong 
interactions. 

These observations suggest the 
following questions for further investigation:
(i) Does dimensional scaling also hold for other Hamiltonians?
(ii) Is there a way to systematically improve on the dimensional
scaling expression, reducing the remaining difference to the
benchmark data without fitting or empirical input?
(iii) Does dimensional scaling hold for other quantities than energies?
A partial answer to question (i) is given by Eq.~(\ref{virial}), which shows
that dimensional scaling can also be expected for the electron liquids.
A related, but not identical, type of dimensional scaling was recently 
also observed for the Heisenberg model.\cite{hemo} Preliminary work on the 
negative $U$ (attractive) Hubbard model also indicates the existence of
simple scaling laws.\cite{nonlinear} The present approach is
expected to be particularly useful for model Hamiltonians that are not
Bethe-Ansatz integrable, but can be solved in the 2-site limit, such as
the periodic Anderson model.
As for question (ii), the answer is `yes'. A simple nonempirical
way to improve on virial-based dimensional scaling will be presented in
a forthcoming publication.\cite{nonlinear}

{\it Acknowledgments}
This work was supported by FAPESP and CNPq. We thank V.~L.~L\'{\i}bero, 
V.~L.~Campo and H.~J.~P.~Freire for useful discussions.

\end{document}